\documentstyle[preprint,aps]{revtex}
\begin{document}
\draft   

\title{Magnetohydrodynamic Equilibria of a Cylindrical Plasma \\with
Poloidal Mass Flow and Arbitrary Cross Section Shape}

\author{ G. N. Throumoulopoulos\thanks{{\it e-mail:} gthroum@cc.uoi.gr}
and G. Pantis\thanks{{\it e-mail}: gpantis@cc.uoi.gr 
\newline \newline
{\sf To be published in Plasma Physics and Controlled Fusion}
 }}
\address{Theoretical Physics Section,\\ Physics Department, University of
Ioannina,\\
 GR 451 10, Ioannina, Greece}
\maketitle
\begin{abstract}
The equilibrium of a cylindrical plasma with  purely poloidal mass flow
and cross section of arbitrary shape is investigated within the framework of
the ideal MHD theory. For the system under consideration it is shown that 
only incompressible  
flows are possible and, conscequently, the general two dimensional flow 
equilibrium equations reduce to a single second-order quasilinear  partial differential
equation for the poloidal magnetic flux function 
$\psi$,
in which  four profile functionals of $\psi$ 
 appear.  Apart from a singularity occuring 
when the modulus of Mach number 
associated with the Alfv\'en velocity for the poloidal magnetic field is
unity, this equation is always elliptic and  permits the construction of
several classes of analytic solutions. Specific exact equlibria for a plasma confined
within  a perfectly conducting circular  cylindrical boundary and  having i) a flat
current density  and ii) a peaked current density are obtained and studied. 
\end{abstract}
\narrowtext

\section{ Introduction}  

Plasma rotation of the order of sound speed has been
observed in many tokamaks heated by neutral beams, such
as JET \cite{CoBe} and TFTR \cite{ScBi}. Early reported data
relates to purely toroidal rotation \cite{Isl}; more recent
measurements show that poloidal rotation can occur as
well \cite{Ta}.  It has also been found that
poloidal flows in the outer region of tokamaks driven by
strong electric fields play an important role in the L-H
transition (e.g. \cite{IdHi,Ma}). 

Although theoretical studies of flowing plasmas began
in the mid fifties \cite{Wo}-\cite{FrRo}, since the early eighties
there has been  an
increasing interest (motivated from the above mentioned observations)
  mainly in the investigation of
equilibrium and stability of plasmas with mass flow
\cite{ZeGr}-\cite{ChMo}.
In particular, the symmetric equilibrium states  of a
 plasma with mass flow in a two dimensional geometry are
governed by a second-order partial semilinear differential equation for the
poloidal magnetic flux function $\psi$, which contains five
free surface quantities (i.e. quantities solely dependent on $\psi$)
 in
conjuction with a nonlinear algebraic Bernouli equation (e.g. 
\cite{ZeGr}-\cite{ZeSt}).
 Unlike in static equilibria, the above
mentioned   differential equation is not always
elliptic; there are three critical values of  the modulus of Mach number 
$|M|$ of the  {\em poloidal} flow with respect to the
poloidal magnetic field, at which the type of this equation
changes, i.e. it becomes alternatively elliptic and hyperbolic.
  To solve the equilibrium problem in
the two elliptic regions, several computer codes have been
developed (e.g. \cite{SeGr,KeTo,ZeSt}).
For purely toroidal flow
the partial differential equation becomes elliptic and
can be solved analytically; analytic solutions were obtained
in Refs. \cite{MaPe80} -\cite{ThPa}.  Owing to the
complexity  of the  equilibrium equations, 
 the construction of
analytic solutions in the case of arbitrary flows 
(toroidal and poloidal) is  difficult. To our best knowledge
the only analytic solution of this kind is that obtained by
Mashke and Perrin \cite{MaPe84} in the specific case of small
ratio of poloidal to toroidal magnetic field and small $\beta$ defined as the
ratio of the kinetic energy to the magnetic energy.

In the present  paper we study the equilibrium of a
cylindrical plasma with  purely poloidal mass flow 
in the framework of ideal MHD theory.  Several physics aspects of a similar system has
been often considered in the literature, e.g. i)  the numerical study of Ref.
\cite{Bi}  concerning the current sheet formation in a cylindrical plasma with
rectangular cross section and purely poloidal flow and ii) the analytic
study of Ref. \cite{ChMo} on nonlinear interactions of tearing modes
 in plane geometry in the presence of sheared poloidal flow. The
plasma cross section of the system under consideration is arbitrary  and
therefore the  configuration  is two-dimensional.  No restriction is
 imposed either on  the components of the magnetic field
or  $\beta$.  We show  that only incompressible equilibrium flows are possible
and that  several classes of analytic equilibrium solutions can be obtained.

 The equilibrium equations of the system under
consideration are presented in Section II. In Section III we derive exact 
equilibrium solutions.
Our conclusions are summarized in Section IV.
\section{ Equilibrium equations}
The system under consideration is a
cylindrical  plasma with  
mass flow and arbitrary cross section shape.  For this
configuration  convenient coordinates are
$\xi$, $\eta$ and $z$ with unit basis vectors $ {\bf e}_\xi$, 
${\bf e}_\eta$,  ${\bf  e}_z$, where  ${\bf  e}_z$
is parallel to the axis of symmetry and $\xi$, $\eta$
are generalized coordinates pertaining to the
poloidal cross section.
 Specific examples
are circular  and elliptical cylindrical coordinates.
The equilibrium quantities do not depend on $z$.
The equations governing the equilibrium states can be obtained from
the general two dimensional equilibrium equations with flow, e.g.
Refs. \cite{MoSo}, \cite{AgTa}, \cite{SeGr}. In particular, the components of
the momentum conservation equation
$$
\rho({\bf v}\cdot\nabla){\bf v} = {\bf j}\times {\bf B} - \nabla P
$$
along the magnetic field and perpendicular to a magnetic surface,
respectively, are put in the form 
\begin{equation}
{\bf B} \cdot \left[{\bf \nabla} \left(\frac{v_z^2}{2}
+ \frac{\left|\nabla F\right|^2}{2\rho^2}\right)
+ \frac{\nabla P}{\rho}\right] = 0 
                                               \label{R1}
 \end{equation}
and
\begin{eqnarray} 
{\bf \nabla} \cdot 
\left[\left(1- \frac{(F^\prime)^2}{\rho}\right)
{\bf \nabla}\psi\right]
+ \frac{F^{\prime\prime}F^\prime |\nabla\psi|^2}{\rho}
+ \frac{B_z\nabla B_z\cdot\nabla \psi}
     {|\nabla \psi|^2}  & &  \nonumber \\
+ \rho \frac{{\bf \nabla} \psi}{|{\bf \nabla} \psi|^2 }\cdot
\left[{\bf \nabla} \left(\frac{(F^{\prime})^2 B^2}{2\rho^2}\right)
+\frac{{\bf \nabla} P}{\rho}\right]=0.                                      						    
                                           \label{R2} 
\end{eqnarray}
Here, $\psi(\xi,\eta)$,
$F(\xi,\eta)$, $B_z(\xi,\eta)$ and $v_z(\xi,\eta)$  are stream functions in terms of
wich  the divergence free fields, i.e. the magnetic field ${\bf B}$,
the current density ${\bf j}$ and the mass flow $\rho {\bf v}$, are expressed as
\begin{equation}
{\bf B}=B_z{\bf e}_z + {\bf e}_z\times\nabla\psi,
                                            \label{R3}
\end{equation}
\begin{equation}
{\bf j}=\nabla^2\psi{\bf e}_z  - {\bf e}_z\times\nabla B_z
                                             \label{R4}
\end{equation}
and
\begin{equation}
\rho{\bf v}=\rho v_z{\bf e}_z + {\bf e}_z\times \nabla F;
                                               \label{R5}
\end{equation}
the function $F$  and the  electrostatic potential $\Phi$ 
are surface quantities satisfying the equations
\begin{equation}
B_z -F^\prime(\psi) v_z= X(\psi)
                                                        \label{R6}
\end{equation}
($X$ also is a surface quantity.)
and
\begin{equation}
\frac{B_z}{\rho}F^\prime(\psi) - v_z = \Phi^\prime(\psi);
\vspace{1cm}
                                                        \label{R7}
\end{equation}
the prime denotes derivative with respect to $\psi$; SI units are used and the vacum magnetic
permeability $\mu_0$ is set to unity. It is noted here that a particular
equation of state, e.g. either isentropic or isothermal magnetic surfaces,
has not been employed  in order to
express in Eq. (\ref{R2})  the pressure
gradient $P^\prime$ in terms of surface quantities; thus, Eq. (\ref{R2}) holds for any
equation of state.

We now restrict  attention to purely poloidal flow and therefore 
set $v_z=0$.
 Then, Eq. (\ref{R6}) implies that
the function $B_z$ is a surface quantity, i.e.
$B_z=B_z (\psi)$.
From equation (\ref{R7}) it follows then that 
\begin{equation}
\rho = \rho (\psi),																													 \label{R8}
 \end{equation}
and therefore the mass density and the magnetic  field share also the
same flux surfaces.  Using Eq. (\ref{R8}), the mass
conservation equation ${\nabla}\cdot \left(\rho {\bf v}\right)=0$ yields
$
{\bf \nabla} \cdot {\bf v} = 0$.												 
Thus, it turns out  that for a cylindrical
plasma and purely poloidal mass flow
only incompressible equilibrium flows are
possible.

With the use
of Eq. (\ref{R8}), Eq. (\ref{R1}) can now be readily integrated, yielding
an expression for the pressure, i.e.
 \begin{equation} 
P = P_s
(\psi) - \frac{F'^2}{2\rho} |{\bf \nabla} \psi|^2.
																																															\label{21}
 \end{equation}
We note here that, unlike in static 
equilibria, in the presence of flow  magnetic surfaces do not coincide with
isobaric surfaces  because the momentum
balance equation implies that ${\bf B} \cdot {\bf \nabla} P$ in
general differs from zero.
In this respect, the term $P_s(\psi)$ is the static part of the pressure which
does not vanish  when $F^\prime$ is set to zero; thus, the second term makes the
isobaric  surfaces to detach from magnetic surfaces.

Before proceeding to obtain the final form of the equilibrium equation we examine a
particular case, i.e. when the   functions $F$ and $\rho$ satisfy the relation
\begin{equation}
\frac{\left(F^\prime \right)^2}{\rho}=1. 
                                                  \label{21a}
\end{equation}
Then,  the second-order
differential operator in Eq. (\ref{R2}) is singular and
since $F$ and $\rho$ are surface quantities 
 this singularity occurs uniformly over a
magnetic surface. 
On the basis of Eq. (\ref{R5}) for $\rho {\bf v}$ and the
definitions $v_{Ap}^2\equiv\frac{\textstyle |\nabla \psi|^2}
{\textstyle \rho}$ for the Alfv\'en velocity associated with the
poloidal magnetic field and the Mach number $M^2\equiv\frac{\textstyle v^2}
{\textstyle v_{Ap}^2}$, Eq. (\ref{21a}) can be written as
$
M^2= 1.
$
Assuming now $\frac{\textstyle (F^\prime)^2}{\textstyle \rho}\neq 1$,
inserting Eq. (\ref{21}) into Eq. (\ref{R2}) and setting $v_z=0$,
 Eq. (\ref{R2})
 after some straightforward algebra, reduces to
the {\em elliptic} differential equation
\begin{equation} 
\left[1 - \frac{(F^\prime)^2}{\rho}\right] {\bf \nabla}^2 \psi +
\frac{F^\prime}{\rho} \left(\frac{F^\prime}{2} 
 \frac{\rho^\prime}{\rho}
-F^{\prime\prime}\right) |{\bf \nabla} \psi|^2 
  +  \left(P_s +\frac{ B_z^2}{2}\right)^\prime = 0.
                                        			    \label{23}
\end{equation}
The absence of any hyperbolic regime in Eq. (\ref{23})
can be understood by noting
that, as is well known from the gass dynamics, the flow must be
compressible to allow the governing partial differential
equation to depart from ellipticity. It may also be noted that in
axisymmetric equilibria with arbitrary flows,  e.g. Eq. (\ref{R2}), $|M|=1$ is not
a transition point. Finally, we note that
replacing the stream functions by the corresponding field
quantities, Eq. (\ref{23}) is put in the form 
\begin{equation}
\frac{d}{dr} \left(P +\frac{|{\bf B}|^2}{2}\right) +
\frac{B^2_\theta}{r} - \rho \frac{v^2}{r} = 0, 
                                   							  \label{21c}
 \end{equation}
the terms of which have a readily unterstandable physical meaning.

\section{Exact Solutions}

In this section, Eq. (\ref{23}) is solved for a specific
choice of the flux functions $\rho(\psi)$ and
$F(\psi)$, i.e. when these functions satisfy the relation
\begin{equation}
\frac{\rho'}{\rho} = 2 \frac{F^{\prime\prime}}{F^\prime}.
																																													\label{28}
 \end{equation}
This ansatz makes the second term in Eq. (\ref{23}) to vanish.
Eq. (\ref{28}) is an ordinary differential equation for
$\psi$, which can easily be integrated.  Its solution is
\begin{equation}
 \frac{(F^\prime)^2}{\rho}=M_c^2,
																																															\label{29}
 \end{equation}
where $M_c^2=$const. $\neq 1$. The choice (\ref{28}) leads
therefore to flows with constant Mach numbers. Solution
(\ref{29}) makes the factor multiplying the second-order
differantial operator in Eq. (\ref{23}) constant; thus, Eq. (\ref{23})
reduces to
\begin{equation} 
{\bf \nabla}^2 \psi + 
\frac{1}{1-M_c^2} \left( P_s + \frac{B_z^2}{2}\right)^\prime = 0.
																																															\label{30}
 \end{equation}
This  is similar in form to the equation
governing static equilibria;  the only explicit reminiscent of
 flow is the presence of $M_c$. Eq. (\ref{30}), which is still
a semilinear  partial differential equation, can be linearized
 for several choices of $P_s + \frac{\textstyle B_z^2}{\textstyle 2}$,
 e.g. a)
for $P_s + \frac{\textstyle B_z^2}{\textstyle 2}$ linear in $\psi$  becomes an
inhomogeneous Laplace equation with constant inhomogeneouity 
and b) for $P_s + \frac{\textstyle B_z^2}{\textstyle 2}$
 quadratic in $\psi$ 
 becomes i) a homogeneous
Helmholtz equation for $M_c^2 > 1$ and ii) a homogeneous diffusion equation 
for
$0< M_c^2 <1$. Several classes of analytic solutions can
therefore be
derived. In the following, specific exact equilibria will be
constructed for a circular cylindrical plasma having i) a flat
current density profile and ii) a  peaked current density
profile. 
%
%
%

\subsection{Flat Current Density Profile}

To describe a plasma which is confined within a circular
infinitely conducting cylinder of radius $r_0$, the coordinates
$\xi$, $\eta$, $z$ are specified to be the circular cylindrical coordinates
$r$, $\theta$, $z$ with unit basis vectors ${\bf  e}_r$, ${\bf
e}_\theta$, ${\bf e}_z$.  The equilibrium quantities
depend just on $r$.  
To obtain  equilibria having a flat current density profile, we employ the
ansatz  $\left(P_s + \frac{\textstyle B_z^2}{\textstyle 2}\right)^\prime= $ 
const.  The simplest non trivial  solution of Eq. (\ref{30})
is written then as
\begin{equation}
\psi \propto  \tau^2,
																																															\label{31}
 \end{equation}
where $\tau \equiv \frac{\textstyle r}{\textstyle r_0}$.  
Solution (\ref{31}) yields a singly peaked parabolic pressure profile
\begin{equation}
P (\rho) = P (0) (1-\tau^2),
																																															\label{32}
 \end{equation}
a flat ``toroidal" current density
\begin{equation}
j_z = -\left(\frac{\alpha}{1-M_c^2}\right) 
       \frac{\sqrt{P(0)}}{r_0},  
																																																\label{35}
 \end{equation}
\begin{equation}
B_z(\rho) =\left\{ B^2_z(0) + 2P(0)
\left[1-\left(\frac{\alpha}{1-M_c^2}\right)^2\right]
\tau^2\right\}^{1/2},
																																														 \label{33}
 \end{equation}
\begin{equation}
B_\theta(\rho) = -\left(\frac{\alpha}{1-M_c^2}\right) \sqrt{P(0)}
\tau,
																																																\label{34}
 \end{equation}
\begin{equation}
{\bf v} =\frac{F^\prime}{\rho}\left({\bf e}_z\times\nabla\psi\right)
=\frac{\pm |M_c|}{1-M_c^2}
\alpha\sqrt{P(0)} \frac{1}{\sqrt{\rho}}{\bf e}_\theta
																																															  \label{36}
 \end{equation}
and
\begin{equation}
{\bf E} = -B_z\frac{F^\prime}{\rho}\nabla\psi=\frac{\mp |M_c|}{1-M_c^2} \alpha
          \sqrt{P(0)}\frac{B_z(\rho)}{\sqrt{\rho}} {\bf e}_r.
																																															  \label{37}
 \end{equation}
Here, $\alpha$ is a parameter which, together with 
$M_c$, describes the magnetic
properties of the plasma, i.e. the plasma is diamagnetic for
$\left(\frac{\textstyle\alpha}{\textstyle 1-M_c^2}\right)^2<1$
and paramagnetic for $\left(\frac{\textstyle
\alpha}{\textstyle 1-M_c^2}\right)^2>1$.  For vanishing flow ($ M_c=0$) the
equilibrium reduces to the exact  axisymmetric equlibrium solution, in the
limit of infinite  aspect ratio, which was suggested by Shafranov \cite{Sha} and
has been extensively used by Solov'ev \cite{Sol}. In Eqs.
(\ref{36}) and (\ref{37}), the density $\rho$ remains a free function of
$\tau$ and therefore a variety of velocity  and electric field
profiles can be derived. As an example, we consider an equilibrium having
a singly  peaked
density profile:
 \begin{equation} \rho =
\frac{\rho(0)}{k^\nu} (k-\rho^2)^\nu,  
                            																			 \label{38}
 \end{equation}
where $\nu >0$ and $k>1$. The parameter
$k$ here is necessary in order to make $\rho$ non-vanishing on
plasma surface ($\rho=1$). (Note  that $\rho$
appears   in the denominator  of Eqs. (\ref{36}) and
(\ref{37})). This agrees qualitatively with experimental
results \cite{Ma,BuGo}, according to which finite densities and very 
low temperatures were measured
in the edge region of plasmas with  mass flow. For the
density profile (\ref{38}), the velocity profile becomes hollow
and the temperature profile, consistent with Eq. (\ref{32}), is 
given by
\begin{equation}
T= T(0) k^\nu \frac{(1-\rho^2)}{(k-\rho^2)^\nu}. 
                   																													 \label{39}
\end{equation}

\subsection{ Peaked current density profile}
 
For simplicity we ignove here  diamagnetic effects, viz. we
consider $\beta_p=1$ equilibria. (The construction of $\beta_p
\neq 1$ equilibria is,  however, possible).  Accordingly, for
$B_z = B_0=$ const. and 
$$P_s\alpha (1-M_c^2) \psi^2 $$ 
the simplest solution of
Eq. (\ref{30}) is
\begin{equation}
\psi = \psi(0) J_0 (\zeta).
																																															  \label{40}
 \end{equation}
$J_0$ and $J_1$ below are 
 zeroth- and first-order
Bessel functions, respectively;  $\zeta\equiv
\frac{\textstyle \lambda_0r_0}{\textstyle r}$, where $\lambda_0$
is the first zero of $J_0$. Solution (\ref{40}) yields a
peaked pressure profile 
\begin{equation}
P = P_s(0) \left[(1-M_c^2) J_0 (\zeta) + M_c^2
    \left(J^2_1 (\lambda_0)
- J^2_1 (\zeta)\right)\right],
																																															  \label{41}
 \end{equation}
a singly peaked current density profile
\begin{equation}
J_z = J_z(0) J_0 (\zeta),
																																																  \label{43}
 \end{equation}
\begin{equation}
B_\theta = - \frac{\lambda_0}{r_0} \psi(0) J_1 (\zeta), 
																																																	\label{42}
 \end{equation}
\begin{equation}
 {\bf v} = \pm |M_c| \frac{\lambda_0 \psi(0)}{r_0} \frac
{J_1 (\zeta)}{\sqrt{\rho}} {\bf e}_\theta
																																																 \label{44}
 \end{equation}
and
\begin{equation}
 {\bf E} = \mp |M_c| \frac{\lambda_0 \psi(0) B_0}{r_0}
\frac {J_1 (\zeta)}{\sqrt{\rho}} {\bf e}_r.
																																														  \label{45}
 \end{equation}
Here, $P_s(0)=P(0)|_{M_c=0}$ is the static pressure at the plasma center.
The above equilibrium has the following characteristics:
\begin{enumerate}
\item Both the pressure and the current density profiles vanish
on the plasma surface.
\item The magnetic field is independent of flow.
\end{enumerate}
 We note that the requirement $P(0)>0$ imposes a restriction on the
values of Mach number, i.e. $M_c^2 < \left[1-J_1
(\lambda_0)^2\right]^{-1}$.
By inspection of Eq. (\ref{41}) it turns out that
as flow increases, i.e. when
$|M_c|$ takes higher values, the pressure  P takes lower values 
and its profile
becomes flatter;  the negative term
$\frac{\textstyle dP}{\textstyle dr}$ 
takes therefore  higher values.  On
the other hand, as  $|{\bf v}|$ increases, the flow term
$-\rho \frac{\textstyle v^2}{\textstyle r}$ (for a given density profile)
 becomes more negative. Thus, the quantity $\frac{\textstyle
dP}{\textstyle dr} - \rho \frac{\textstyle v^2}{\textstyle r}$
becomes flow independent and therefore guarantees the  equilibrium
equation (\ref{21c}), in which the magnetic field terms (for the solution 
(\ref{40})) are independent of
flow.

\section{ Conclusions}

Using the general two dimensional ideal MHD equlibrium  equations with flow,
 we investigated 
the stationary states  of a cylindrical plasma  with purely poloidal flow and arbitrary
cross-section shape and showed that the flows are necessarily incompressible.
 Consequently: 
(i)  The pressure
 profile is expressed in terms of a static pressure surface quantity $P_s$ and
a flow dependent term not
remaining constant on a magnetic surface (Eq. (\ref{21})),
 which is consistent with the
fact  that for equilibria with mass flow the magnetic surfaces are detached from the
isobaric surfaces.
(ii) The flow equations reduce 
to a single second-order quasilinear partial differential equation for 
$\psi$, (Eq. (\ref{23})),  which contains four  surface quantities, i.e.
$P_s(\psi)$, the flow function 
$F(\psi)$, the ``toroidal" magnetic field $B_z(\psi)$ and the density
 $\rho(\psi)$.
 One has to specify these quantities together with the boundary conditions for
$\psi$ in order to determine the equilibrium completely.  A singularity occurs in the
differential equation  when the modulus of the Mach number $|M|$, associated with the
Alfv\'en velocity for the poloidal magnetic field, is unity.  
 Apart from
this case, this equation is always
elliptic. This  is consistent with the fact that the existence of
hyperbolic regimes in the governing equilibrium differential equation is
possible only in compressible fluids.   For velocities
 satisfying the relation $|M|= \text{const.} \neq 1$, the equilibrium equation 
 takes a simpler form (Eq. (\ref{30})), which is amendable to 
 several classes of analytic solutions. We constructed and studied 
particular exact equilibria for a plasma confined within a
circular cylindrical perfectly conducting boundary  having i) a flat
current density profile and ii) a peaked current density.

The analytic equilibria with poloidal flow we constucted in the present paper can be
the basis of stability and transport investigations, which would be of relevance to magnetic
confinement systems. In particular,  they
 may help in
understanding the physics involved in the transition from the low confinement mode to the high
confinement mode in tokamaks, which is crucially related to poloidal flow. The derivation of new
analytic  solutions in more realistic situations, e.g. axisymmetric equilibria, hellicaly
symmetric equilibria (even in specific cases such as in rotating plasmas  with  a single
velocity component  or/and small $\beta$), might help to easily get physical insight of plasmas
with flows. Solutions of this kind might also be used for checking relevant equilibrium
codes. %
\widetext


\begin{references} 
\bibitem{CoBe} Core  W. G. F.  van Bell P. and  Sadler G. 1987 
                   {\it  Proc. of the 14st European Conf. Controlled Fusion and
                  Plasma Physics} (Madrid 1987) vol 11D  p 49   
                    (Geneva: European Physical Society)
                   
\bibitem{ScBi}  Scott S. D. {\it et al.} 1987
                   {\it Proc. of the 14st European Conf. Controlled Fusion and Plasma
                  Physics}  (Madrid 1987) vol 11D p 65
                  (Geneva:  European  Physical Society)
                  
\bibitem{Isl}  Isler  R. C. {\it et al.} 1983 {\it Nucl. Fusion} {\bf 27} 1017 
\bibitem{Ta}  Taylor R.  {\it et al.}
                   {\it Proc.  13th IAEA Conf. on Plasma Physics and
                  Controlled Nuclear Fusion Research}  (Washington 1990)
                  paper CN-53/A-6-5
                  (Viena: IAEA) 
\bibitem{IdHi}  Ida K. and  Hidekuma S. 1990 {\it Phys. Rev. Lett.} {\bf 65} 1364  
\bibitem{Ma} Matsumoto H. {\it et al.} 1992 {\it Plasma Phys. Contr. Fusion} {\bf 34}
              615
\bibitem{Wo}   Woltjer  L. 1959 {\it Astrophys.} {\bf 130} 405 
\bibitem{Gr}  Grad H. 1960 {\it Rev. Mod. Phys.} {\bf 32} 830 
\bibitem{FrRo} Frieman  E. and  Rotenberg M. 1960 {\it Rev. Mod. Phys.} {\bf 32} 898
\bibitem{ZeGr} Zehrfeld H. P. and  Green B. J. 1972 {\it Nuclear Fusion} {\bf 12}
               569 
\bibitem{MoSo}  Morozov A. I. and  Solov'ev L. S. 1980 
               {\it Reviews of Plasma
               Physics}  vol 8  p 18 (edited by  M. A. Leontovich, Consultans Bureau
               New York)
\bibitem{AgTa}  Agim Y. Z. and  Tataronis J. A. 1985 {\it J. Plasma Phys.} {\bf 34} 337
\bibitem{Ha79} Hameiri E. 1979  {\it Phys. Fluids} {\bf 22} 89  
\bibitem{SeGr}  Baransky Y. A 1987 {\it Ph.D. Thesis} Department of Applied
               Physics Columbia University 
\bibitem{SeGr} Semenzato S. Gruber R. and Zehrfeld H. P. 1984
               {\it Comput. Phys. Rep.} {\bf 1} 389 
\bibitem{KeTo}  Kerner W. and  Tokuda S. 1987 {\it Z. Naturforsch.} {\bf 42a}
               1154 
\bibitem{ZeSt} R. \.Zelazny {\it et al.},
               {\it Plasma Phys. Contr. Fusion} {\bf 35}, 1215 (1993).
\bibitem{AgBh}  Agarwal A.  Bhattacharyya K. S. N. and  Sen A 1992 {\it Plasma Phys.
               and Contr. Fusion} {\bf 34} 1211
\bibitem{BoIa}  Bondeson A. Iacono  R. and  Bhattacharjee A. 1987 {\it Phys. Fluids}
              {\bf 30} 2167 
\bibitem{BhAv}   Bhattacharyya S. N. Avinash  K. and  Sen A.  1992 {\it Plasma Phys.
               and Contr. Fusion} {\bf 34} 457 
\bibitem{Bi}   Biscamp D. 1993 {\it  Phys. Fluids B} {\bf 5} 3893 
\bibitem{ChMo}  Chen X. L. and  Morrison P. J. 1992  {\it Phys. Fluids B} {\bf 4} 845 
\bibitem{MaPe80} Maschke  E. K.  and  Perrin H. 1980 {\it Plasma Phys.} {\bf
                22} 579 
\bibitem{ClFa}  Clemente R. A. and  Farengo R. 1984 {\it Phys. Fluids}
                {\bf 27}, 776 
\bibitem{ThPa}  Throumoulopoulos G. N. and  Pantis G. 1989 {\it Phys. Fluids B} {\bf 1} 1827 
\bibitem{MaPe84}  Maschke E. K. and  Perrin H. J. 1984 {\it Phys. Lett}  {\bf 102A} 106
\bibitem{BuGo} Bugarya  V. I.  {\it et al.} 1983 {\it Pis'ma Zh. Eksp. Teor. Fiz.} {\bf
              38}
               337  [Transl 1983 {\it JETP Lett.} {\bf 38} 404]
\bibitem{Sha}  Shafranov V. D. 1966 {\it Reviews of Plasma Physics} vol 8
              (edited by M. A. Leontovich Consultans Bureau
               New York) 
\bibitem{Sol}  Solov'ev L. S.  1976 {\it Reviews of Plasma Physics} vol 5
                (edited by M. A. Leontovich  Consultans Bureau
               New York) 
%
\end{references}
\end{document}